\documentclass[twocolumn,aps,showpacs]{revtex4-1}

\usepackage{amsfonts}
\usepackage{amsmath}
\usepackage{graphicx}
\usepackage{bm}
\usepackage{amssymb}
\usepackage{dcolumn}
\usepackage{color,ulem}
\usepackage{epstopdf}

\begin{document}

\title{Analysis of the spin Hall effect in CuIr alloys: Combined approach of density functional theory and Hartree-Fock approximation}

\author{Zhuo Xu$^{1,2}$,
Bo Gu$^{1,2}$, Michiyasu Mori$^{1,2}$, Timothy Ziman$^{3,4}$, and Sadamichi Maekawa$^{1,2}$}

\affiliation{$^{1}$Advanced Science Research Center, Japan Atomic Energy Agency, Tokai 319-1195, Japan
\\ $^{2}$CREST, Japan Science and Technology Agency, Sanbancho, Tokyo 102-0075, Japan
\\ $^{3}$Institut Laue Langevin, 71 avenue des Martys, CS-20156, F-38042 Grenoble Cedex 9, France
\\ $^{4}$LPMMC (UMR 5493),CNRS and  Universit\'{e} Joseph-Fourier Grenoble, 38042 Grenoble, France
}

\date{\today}
\begin{abstract}
We analyze the spin Hall effect in CuIr alloys in theory by the combined approach of
the density functional theory (DFT) and Hartree-Fock (HF) approximation.
The SHA is obtained to be negative without the local correlation effects.
After including the local correlation effects of the $5d$ orbitals of Ir impurities,
the SHA becomes positive with realistic correlation parameters, and
consistent with experiment [Niimi et al., Phys. Rev. Lett. {\bf 106}, 126601 (2011)].
Moreover, our analysis shows that the DFT+HF approach is a convenient and general method to study the
influence of  local correlation effects on the spin Hall effect.
\end{abstract}

\pacs{71.70.Ej, 72.25.Ba, 85.75.-d}
\maketitle

\section{Introduction}

The spin Hall effect (SHE) converts charge current into spin current, which is crucial for the further
development of spintronic devices. The key material parameter in a device based on this effect is the spin Hall angle
(SHA): the ratio between the induced spin Hall current and the input charge current. If the sign of SHA
changes, the direction of the induced spin Hall current is also reversed.
In the experiment  on CuIr alloys,
the dominant contribution to the SHE was verified to be  by an extrinsic skew scattering mechanism
and the SHA was measured to be positive 2.1\% \cite{Niimi}.

According to  the skew scattering approach of Fert and Levy \cite{Fert,Fert-Levy},
the phase shift of the $6p$ orbitals of Ir
is decisive as to the sign of the SHA of CuIr.
A positive SHA can be obtained as long as the phase shift of $6p$
has  a small positive value \cite{Fert-Levy}, and our aim is to
find  a microscopic argument for this, rather than treating it as a free parameter.
Other approaches based on the Boltzmann equation and Kubo-Streda formula
even give negative values of SHA, (according to  the definition of SHA from resistivities)\cite{Fedorov},
opposite to the experimental sign.
Therefore, a clear and convenient theoretical approach which can
reproduce and explain the  sign of the SHA is still required.

In the present work, by the combined approach of the density functional theory (DFT) and
Hartree-Fock (HF) approximation,
we calculate the SHA including  correlation effects,
and find  that the local correlation effects of the $5d$ orbitals of Ir
give the  sign of SHA consistent with experiment.

\section{Skew Scattering}

For the CuIr alloys, the spin orbit interactions (SOI) in the $5d$ orbitals
of the Ir impurities induce the extrinsic SHE.
It has been observed in experiment that
the spin Hall resistivity increases linearly with the impurity concentration,
so that the SHE is predominantly attributed to a
skew scattering extrinsic contribution \cite{Niimi}.
Thus, the nonmagnetic CuIr alloys can be described by
a single-impurity multi-orbital Anderson model \cite{Anderson}:
\begin{equation}
\begin{split}
H_{00}=&\sum_{\textbf{k},\alpha,\sigma}\epsilon_{\alpha\textbf{k}}
  c^{\dag}_{\textbf{k}\alpha\sigma}c_{\textbf{k}\alpha\sigma},\\
H_{0}=&H_{00}
   +\sum_{\textbf{k},\alpha,\beta,\sigma}(V_{\beta\textbf{k}\alpha }
    d^{\dag}_{\beta\sigma} c_{\textbf{k}\alpha\sigma} + \textrm{H.c.})
   + \sum_{\beta,\sigma}\epsilon_{\beta}n_{\beta\sigma},\\
H_{SO}=&\frac{\lambda_{p}}{2}\sum_{\zeta\sigma,\zeta^{\prime}\sigma^{\prime}}d^{\dagger}_{\zeta\sigma}
   (\textbf{l})_{\zeta\zeta^{\prime}}\cdot(\pmb\sigma)_{\sigma\sigma^{\prime}}d_{\zeta^{\prime}\sigma^{\prime}}\\
   &+\frac{\lambda_{d}}{2}\sum_{\xi\sigma,\xi^{\prime}\sigma^{\prime}}d^{\dagger}_{\xi\sigma}
   (\textbf{l})_{\xi\xi^{\prime}}\cdot(\pmb\sigma)_{\sigma\sigma^{\prime}}d_{\xi^{\prime}\sigma^{\prime}},\\
H=&H_{0}+H_{SO}+U\sum_{\xi}n_{\xi\uparrow}n_{\xi\downarrow}\\
  & + \frac{U^{\prime}}{2}\sum_{\xi\neq\xi',\sigma,\sigma^{\prime}}
     n_{\xi\sigma}n_{\xi'\sigma^{\prime}}
   - \frac{J}{2}\sum_{\xi\neq\xi',\sigma}n_{\xi\sigma}n_{\xi'\sigma},
\end{split}
\label{andersonmodel}
\end{equation}
where $\epsilon_{\alpha\textbf{k}}$ is the energy band $\alpha$ of the host Cu,
$\epsilon_{\beta}$ is the energy level of the orbital $\beta$ of the impurity Ir,
and $V_{\beta,\alpha}(\textbf{k})$ is the hybridization between the orbital $\beta$
of Ir and the band $\alpha$ of Cu.
$U$ ($U^{\prime}$) is the on-site Coulomb repulsion within (between) the $5d$ orbitals of Ir,
and $J$ is the Hund coupling between the $5d$ orbitals of Ir.
The relations of $U=U^{\prime}+2J$ and $J/U=0.3$ are kept \cite{Maekawa}.
The SOI is included in both the $6p$ orbitals $\zeta$ and the $5d$ orbitals $\xi$ of Ir,
with the parameters $\lambda_{p}$ and $\lambda_{d}$, respectively.
We include the on-site Coulomb interactions only within the $5d$ orbitals of Ir in Eq.(\ref{andersonmodel}),
but not within the $6p$ orbitals, which are much more extended \cite{Gu-CuBi}.

The SOI included in the $d$ orbitals will split
the $d$ states with the orbital angular momentum $l$ into the states of $d\pm$
with the total angular momentum $j=l\pm\frac{1}{2}$.
The degeneracy of the $d+$ and $d-$ states is six and four, respectively.
For the $5d$ states with SOI of Ir,
we have the relations of
$n_{d+}=N_{d+}^{Ir}/6$ and $n_{d-}=N_{d-}^{Ir}/4$,
where $n_{d\pm}$ are the occupation number of each of the degenerate states $5d\pm$,
and $N_{d\pm}^{Ir}$ are the total occupation number of the $5d\pm$ states.
The values of $n_{d\pm}$ will be between 0 and 1.
The total occupation number of the $5d$ states of Ir $N_{d}^{Ir}=N_{d+}^{Ir}+N_{d-}^{Ir}$.
Similarly, the SOI splits the $p$ orbitals into $p\pm$ states.

Since a net charge cannot exist in metal,
the total occupation numbers of the valence states of $6s$, $6p$ and $5d$ of Ir
is conserved as \cite{Xu}
\begin{equation}
N_{s}^{Ir}+N_{p}^{Ir}+N_{d}^{Ir} = 9,
\label{charge}
\end{equation}
where the occupation numbers are defined via
projections of the occupied states
onto the Wannier states centered at the Ir sites and extended in the whole supercell.

Following the method of Ref. \cite{Guo}
and the definition of SHA $\Theta$ in terms of resistivity $\rho$ \cite{Xu},
the SHA of CuIr can be calculated
from the phase shifts $\delta_{1}^{\pm}$ of the $p\pm$ and $\delta_{2}^{\pm}$ of the $d\pm$
channels as
\begin{equation}
\begin{split}
\Theta & (\delta_{1}^{+},\delta_{1}^{-},\delta_{2}^{+},\delta_{2}^{-}) = A/B,\\
A=& -2[9\sin(\delta^{+}_{1}-\delta^{+}_{2})\sin\delta^{+}_{1}\sin\delta^{+}_{2}\\
&-4\sin(\delta^{+}_{1}-\delta^{-}_{2})\sin\delta^{+}_{1}\sin\delta^{-}_{2}\\
&-5\sin(\delta^{-}_{1}-\delta^{-}_{2})\sin\delta^{-}_{1}\sin\delta^{-}_{2}],\\
B=& 45\sin^2\delta^{+}_{2}+30\sin^2\delta^{-}_{2}+50\sin^2\delta^{+}_{1}+25\sin^2\delta^{-}_{1}\\
&+6\sin\delta^{+}_{1}\sin(2\delta^{+}_{2}-\delta^{+}_{1})+12\sin\delta^{-}_{1}\sin(2\delta^{+}_{2}-\delta^{-}_{1})\\
&+14\sin\delta^{+}_{1}\sin(2\delta^{-}_{2}-\delta^{+}_{1})-2\sin\delta^{-}_{1}\sin(2\delta^{-}_{2}-\delta^{-}_{1}).
\end{split}
\label{SHA0}
\end{equation}
The phase shifts can be obtained by the Friedel sum rule \cite{Fert,Langreth}:
\begin{equation}
\delta_{\mu}^{\pm} = \pi(N_{\mu\pm}^{Ir}-N_{\mu\pm}^{Cu})/D_{\mu\pm},
\label{phaseshifts}
\end{equation}
where $\mu$=1 for $p$ orbitals with the degeneracies $D_{1+}$=4 and $D_{1-}$=2,
$\mu$=2 for $d$ orbitals with $D_{2+}$=6 and $D_{2-}$=4.

\section{DFT Results}

For the DFT calculation, we employ the code of Quantum Espresso (QE) \cite{QE}.
We use a primitive cell of a single Cu atom to calculate the $H_{00}$ exclusively for the case of pure Cu,
and a supercell of Cu$_{26}$Ir to calculate the $H_{0}$ for the case of CuIr alloys.
The cutoff energy of planewaves is 50 Ry. The pseudopotentials are
ultrasoft for calculations without SOI and projector-augmented-wave for calculations with SOI.
The type of exchange-correlation functionals is PBE \cite{PBE}.
The energy convergence limit is 10$^{-8}$ Ry. The $k$ lattice is $8\times8\times8$.

By the DFT calculations of $H_{00}$ and $H_{0}$ in Eq.(\ref{andersonmodel}) for pure Cu and CuIr,
respectively, the occupation numbers of $4s$, $4p$, $3d$ states of Cu and $6s$, $6p$, $5d$ states of Ir
are obtained to be $N_{s}^{Cu}$=0.35, $N_{p}^{Cu}$=0.96, $N_{d}^{Cu}$=9.68 and
$N_{s}^{Ir}$=0.32, $N_{p}^{Ir}$=0.86, $N_{d}^{Ir}$=7.82, respectively.
Thus the total occupation number $N_{s}^{Ir}+N_{p}^{Ir}+N_{d}^{Ir}$=9.0,
confirming the relation in Eq.(\ref{charge}).
Including the SOI, the DFT calculations for pure Cu and CuIr give
the phase shifts of $\delta_{1}^{+}$=-0.09, $\delta_{1}^{-}$=0.06,
$\delta_{2}^{+}$=-0.73 and $\delta_{2}^{-}$=-0.38, by Eq.(\ref{phaseshifts}).
The SHA is obtained to be -2.7\% by Eq.(\ref{SHA0}),
which is close to the prediction in Ref. \cite{Mertig1}, but
is inconsistent with the  positive sign in experiment \cite{Niimi}.

\section{DFT+HF Approach}

\begin{figure}[tbp]
\includegraphics[width = 5 cm]{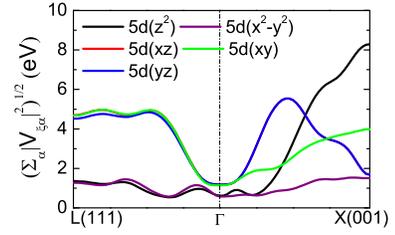}
\caption{The hybridization function between the $5d$ orbitals of the Ir impurity and the bulk Cu host.}
\label{F-hybridization}
\end{figure}

The hybridization between the $5d$ orbitals of the Ir impurity and the Cu host is defined as
\begin{equation}
V_{\xi\textbf{k}\alpha}\equiv \langle\varphi_{\xi}|H_{0}|\Psi_{\alpha}(\textbf{k})\rangle,
\label{hybridization0}
\end{equation}
where $\varphi_{\xi}$ is the Ir impurity state in real space with the $5d$ orbital index $\xi$,
and $\Psi_{\alpha}(\textbf{k})$ is the Cu host state in k-space with the band index $\alpha$ and wavevector $\textbf{k}$.
Following the method in Ref. \cite{GuCon},
and using the post-processor code Wannier90 \cite{wannier90},
$V_{\xi\textbf{k}\alpha}$ were obtained.
In Fig. {\ref{F-hybridization}} we plot the function $(\sum_{\alpha}|V_{\xi\textbf{k}\alpha}|^{2})^{1/2}$.

\begin{figure}[tbp]
\includegraphics[width = 6 cm]{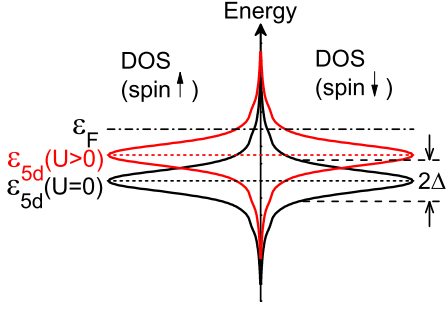}
\caption{Schematic picture of the density of states (DOS)
of the $5d$ virtual bound states with width $\Delta$ of Ir in the nonmagnetic CuIr alloys,
with the on-site Coulomb repulsion $U$=0 (black) and $U>$0 (red), respectively.
$\varepsilon_{5d}$ is the impurity level in the Hartree-Fock approximation
which will increase with $U$, and $\varepsilon_{F}$ is the Fermi level.}
\label{andersonvbs}
\end{figure}

Based on the Anderson model \cite{Anderson},
the $5d$ states of Ir impurities are considered as virtual bound states with width $\Delta$.
Including correlation $U$ on the virtual bound states,
the impurity level increases while the occupation number decreases,
as shown schematically in Fig. {\ref{andersonvbs}}.
Taking the results of $V_{\xi\textbf{k}\alpha}$ by Eq.(\ref{hybridization0}),
the width parameter $\Delta_{\xi}$ of the virtual bound state for each $5d$ orbital $\xi$ of Ir
is obtained by the relation \cite{Anderson}
\begin{equation}
\Delta_{\xi}
=\pi\sum_{\alpha,\textbf{k}}\delta(\epsilon_{F}-\epsilon_{\alpha\textbf{k}})|V_{\xi \textbf{k}\alpha}|^{2},
\label{width}
\end{equation}
where $\epsilon_{F}$ is the Fermi level.
As a result, the width $\Delta$ for the whole $5d$ orbitals of Ir is the average of each $\Delta_{\xi}$,
$\Delta=(\sum_{\xi}\Delta_{\xi})/5$=1.76 eV.

Based on the Anderson model of $H_{0}+H_{SO}$ in Eq.(\ref{andersonmodel}),
for the nonmagnetic CuIr including SOI but without correlation $U$,
there are self-consistent relations between the spin-orbit split states of $5d+$ and $5d-$ of Ir \cite{Anderson}:
\begin{equation}
\Delta \cot(\pi n_{d\pm})=E_{0,d\pm},
\label{sc1}
\end{equation}
where the $5d\pm$ states of Ir under $U$=0 has the energy level of $E_{0,d\pm}$.
The DFT results of $H_{0}+H_{SO}$ in Eq.(\ref{andersonmodel}) give $n_{d+}$=0.734 and $n_{d-}$=0.849.
Then Eq.(\ref{sc1}) gives $E_{0,d+}$=-1.59 eV and $E_{0,d-}$=-3.43 eV.

For correlation $U>0$, based on  Eq.(\ref{andersonmodel})
with HF approximation,
the self-consistent relations are rewritten as
\begin{equation}
\begin{split}
E_{d\pm}=&\Delta \cot(\pi n_{d\pm})\\
=&E_{0,d\pm}+U(\frac{3}{5}n_{d+}+\frac{2}{5}n_{d-})+U^{\prime}(\frac{24}{5}n_{d+}+\frac{16}{5}n_{d-})\\
&-J(\frac{12}{5}n_{d+}+\frac{8}{5}n_{d-}),
\end{split}
\label{sc2}
\end{equation}
from which the $n_{d+}$ and $n_{d-}$ can be obtained for each positive $U$. Eq.(\ref{sc2}) directly includes all the five $5d$ orbitals of Ir,
as well as the local correlations,
and the calculation is self-consistent.

\begin{figure}[tbp]
\includegraphics[width = 8.5 cm]{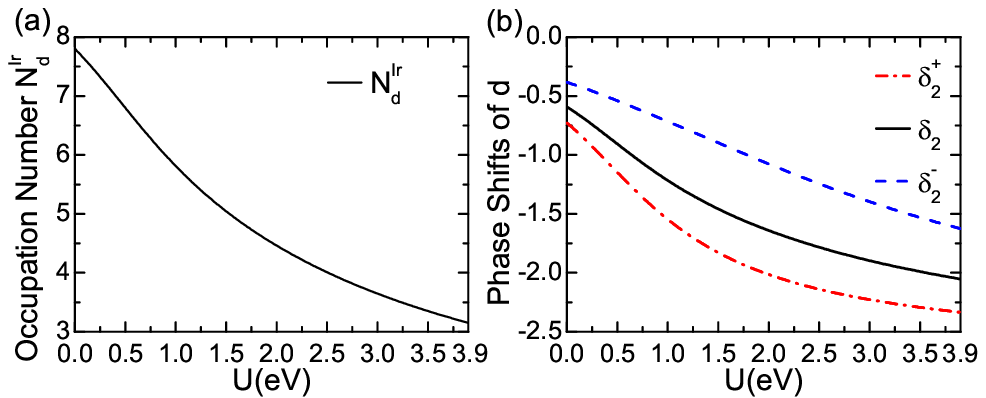}
\includegraphics[width = 8.5 cm]{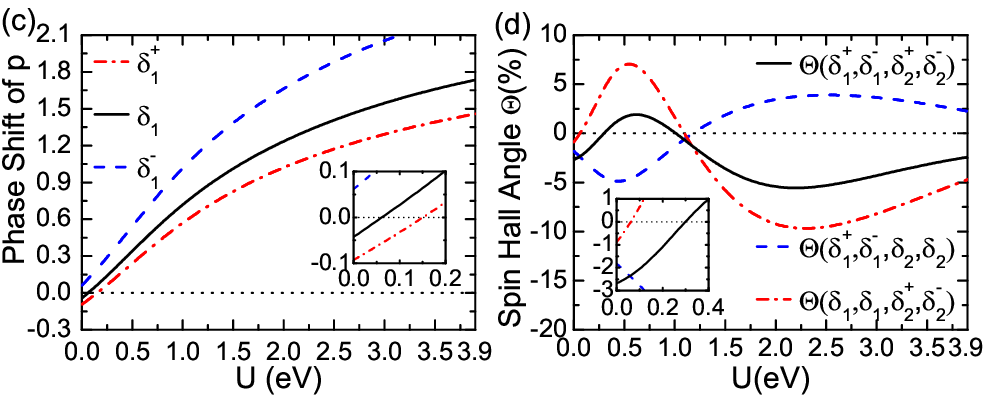}
\caption{(a) The occupation number of the $5d$ orbitals of Ir $N_{d}^{Ir}$,
(b) the phase shifts obtained by Eq.(\ref{phaseshifts}) of
 $d$ orbitals with SOI ($\delta_{2}^{\pm}$)
and without SOI ($\delta_{2}$),
(c) $p$ orbitals with SOI ($\delta_{1}^{\pm}$) and without SOI ($\delta_{1}$) ,
and (d) the SHA $\Theta$ as functions of correlation $U$.
The inserts show the details around $U$=0.}
\label{hf2}
\end{figure}

As $U$ increases from 0,
the occupation number $N_{d}^{Ir}=6n_{d+}+4n_{d-}$ decreases, as shown in Fig. {\ref{hf2}}(a).
The phase shifts $\delta_{2}^{\pm}$ obtained by Eq.(\ref{phaseshifts})
are plotted in Fig. {\ref{hf2}}(b).
The ratios of $N_{p}^{Ir}/N_{s}^{Ir}$=2.7 and $N_{p+}^{Ir}/N_{p-}^{Ir}$=1.4 from DFT
with $U$=0,
and the relation in Eq.(\ref{charge}) are taken to evaluate the occupation numbers
$N_{p(\pm)}^{Ir}$. The phase shifts $\delta_{1}^{\pm}$ obtained
from Eq.(\ref{phaseshifts}) are shown in Fig. {\ref{hf2}}(c).
Finally, the SHA
$\Theta(\delta_{1}^{+},\delta_{1}^{-},\delta_{2}^{+},\delta_{2}^{-})$,
calculated from Eq.(\ref{SHA0}), is shown in Fig. {\ref{hf2}}(d).
In order to compare the contributions from the $p$ and $d$ orbitals separately,
we consider the two limiting cases of $\delta_{1}^{+}=\delta_{1}^{-}=\delta_{1}$
and $\delta_{2}^{+}=\delta_{2}^{-}=\delta_{2}$.
We define $\delta_{1}$ and $\delta_{2}$ from Eq.(\ref{phaseshifts}) with 
the total occupation numbers of each orbital and the degeneracies
6 and 10 respectively,
and plot them in Figs. {\ref{hf2}}(c) and (b).
The SHA of the two limiting cases $\Theta(\delta_{1},\delta_{1},\delta_{2}^{+},\delta_{2}^{-})$ and $\Theta(\delta_{1}^{+},\delta_{1}^{-},\delta_{2},\delta_{2})$ are plotted in Figs. {\ref{hf2}}(d).

In addition, we need to evaluate the range of $U$ within which the nonmagnetic state of CuIr is the ground state.
Following Anderson's method to calculate the critical value of $U$ between nonmagnetic and magnetic states \cite{Anderson},
neglecting the SOI as an approximation, based on the $H-H_{SO}$ in Eq.(\ref{andersonmodel}),
there are self-consistent relations among the five degenerate $5d$ orbitals:
\begin{equation}
\begin{split}
E_{\xi,\sigma}=&\Delta \cot(\pi n_{\xi,\sigma})\\
=&E_{0}+Un_{\xi,-\sigma}+U^{\prime}\sum_{\xi\prime\neq\xi}(n_{\xi\prime,\sigma}+n_{\xi\prime,-\sigma})\\
&-J\sum_{\xi\prime\neq\xi}n_{\xi\prime,\sigma},
\end{split}
\label{sc3}
\end{equation}

For the nonmagnetic case, $n_{\xi,\uparrow}=n_{\xi,\downarrow}=n$,
we have
\begin{equation}
\Delta \cot(\pi n)=E_{0}+Un+8U^{\prime}n-4Jn.
\label{sc4}
\end{equation}
Taking the parameters of $n$=0.78 obtained from the DFT calculation of
$H_{0}$ in Eq.(\ref{andersonmodel}) and $\Delta$=1.76 eV obtained from Eq.(\ref{hybridization0}),
Eq.(\ref{sc4}) gives $E_{0}=\Delta \cot(\pi n)$=-2.16 eV.

For the magnetic case, by differentiating Eq.(\ref{sc3}),
\begin{equation}
\begin{split}
-\frac{\Delta \pi}{\sin^{2}\pi n}\delta n_{\xi,\sigma}=&
U\delta n_{\xi,-\sigma}
+U^{\prime}\sum_{\xi\prime\neq\xi}(\delta n_{\xi\prime,\sigma}+\delta n_{\xi\prime,-\sigma})\\
&-J\sum_{\xi\prime\neq\xi}\delta n_{\xi\prime,\sigma}.
\end{split}
\label{sc5}
\end{equation}
Letting $\delta n_{\sigma}=\sum_{\xi}\delta n_{\xi,\sigma}$,
from Eq.(\ref{sc5})
\begin{equation}
\begin{split}
-\frac{\Delta \pi}{\sin^{2}\pi n}\delta n_{\sigma}
=&U\delta n_{-\sigma}+U^{\prime}(4\delta n_{\sigma}+4\delta n_{-\sigma})-4J\delta n_{\sigma},
\end{split}
\label{sc6}
\end{equation}
\begin{equation}
\frac{\Delta \pi}{\sin^{2}\pi n}(\delta n_{\uparrow}-\delta n_{\downarrow})
=(U+4J)(\delta n_{\uparrow}-\delta n_{\downarrow}).
\label{sc7}
\end{equation}
For the magnetic case, $\delta n_{\uparrow}-\delta n_{\downarrow}\neq0$, thus
\begin{equation}
\frac{\Delta \pi}{\sin^{2}\pi n}=U+4J.
\label{sc8}
\end{equation}

With the parameters of $\Delta$=1.76 eV and $E_{0}$=-2.16 eV already obtained above,
and the fixed relations of $U^{\prime}=U-2J$ and $J=0.3U$ \cite{Xu,Maekawa},
by solving Eqs.(\ref{sc4}) and (\ref{sc8}) simultaneously,
it gives the critical occupation number $n_{c}$ and the critical correlation parameter $U_{c}$
between the nonmagnetic phase and magnetic phase to be $n_{c}$=0.30 and $U_{c}$=3.92 eV.
The critical value of the total occupation number $(N_{d}^{Ir})_{c}=10n_{c}$=3.0.
Thus the results in Fig. {\ref{hf2}} are of the nonmagnetic states.
As the correlation $U$ increases from zero up to the nonmagnetic limit of 3.92 eV,
the SHA is non-monotonic.

\section{Analysis of the Spin Hall Effect}

As shown in Fig. {\ref{hf2}}(a),
as the correlation $U$ on the $5d$ orbitals of Ir increases from 0,
$N_{d}^{Ir}$ decreases,
which is consistent with the picture from the Anderson model \cite{Anderson}
as in Fig. {\ref{andersonvbs}}. Due to the relation in Eq.(\ref{charge}),
the decrease of $N_{d}^{Ir}$ is accompanied by the increase of $N_{p}^{Ir}$.
According to Eq.(\ref{phaseshifts}),
the phase shifts $\delta_{2}$ and $\delta^{\pm}_{2}$ decrease,
while $\delta_{1}$ and $\delta^{+}_{1}$ increase from negative to positive,
and $\delta^{-}_{1}$ is always positive and increases in magnitude,
as shown in Fig. {\ref{hf2}}(b) and (c).

From Fig. {\ref{hf2}}(d), we note that at $U$=0,
the magnitude of the calculated SHA $\Theta(\delta_{1}^{+},\delta_{1}^{-},\delta_{2},\delta_{2})$
with SOI only in the $p$ orbitals is larger than
$\Theta(\delta_{1},\delta_{1},\delta_{2}^{+},\delta_{2}^{-})$,
with SOI only in the $d$ orbitals.
This is consistent with the results in Ref.\cite{Mertig2}.
As the correlation $U$ increases to a realistic value for Ir of around 0.5 eV \cite{UIr},
the SHA including SOI in both $p$ and $d$ orbitals,
$\Theta(\delta_{1}^{+},\delta_{1}^{-},\delta_{2}^{+},\delta_{2}^{-})$,
goes from negative to positive values.
At $U$=0.5 eV, the SHA is +1.6\%,
quite close to the experimental value of +2.1\% \cite{Niimi}.
If the SOI is included only in the $5d$ orbitals of Ir,
the resulting SHA $\Theta(\delta_{1},\delta_{1},\delta_{2}^{+},\delta_{2}^{-})$ still
qualitatively follows the complete function $\Theta(\delta_{1}^{+},\delta_{1}^{-},\delta_{2}^{+},\delta_{2}^{-})$.
If, on the other hand, we consider the SOI only in the $p$ orbitals,
the predicted SHA $\Theta(\delta_{1}^{+},\delta_{1}^{-},\delta_{2},\delta_{2})$ around $U$=0.5 eV is
opposite in sign to the experiment.
This is because the relative magnitudes of $\Theta(\delta_{1},\delta_{1},\delta_{2}^{+},\delta_{2}^{-})$
and $\Theta(\delta_{1}^{+},\delta_{1}^{-},\delta_{2},\delta_{2})$ are reversed as
$U$ increases from 0 to the realistic value.
In addition, the SOI of the more extended $6p$ orbitals of Ir
is likely to be overestimated by the DFT calculation;
thus the contribution due to the SOI in the $p$ orbitals terms may be exaggerated.
These results suggest the physical reason for the SHA of CuIr observed in experiment:
it is the local correlation effects of the $5d$ orbitals of Ir which determine the sign of the SHA.

In conclusion, by the combined approach of DFT and HF approximation,
we show that the local correlation effects of the $5d$ orbitals of Ir
give the  sign of the SHA consistent with  experiment.
This indicates it is a convenient and general method to study the
influence of  local correlations effects on the SHE,
for various combinations of hosts and impurities
and for a  wide range of $U$.

\section*{Acknowledgement}	
This work is supported by Grant-in-Aid for Scientific Research (Grant No.23340093, No.24360036, No.24540387, No.25287094, No.26108716, and No.26247063) and bilateral program from MEXT, by the National Science Foundation under Grant No. NSF PHY11-25915, and a REIMEI project of JAEA.


\end{document}